\documentclass[sigconf]{acmart}

\usepackage{booktabs}
\usepackage{graphicx}
\usepackage{hyperref}
\usepackage{xcolor}

\hypersetup{
    colorlinks=true,
    linkcolor=cyan,
    filecolor=magenta,
    urlcolor=cyan,
}

\setcopyright{rightsretained}

\acmConference[2018]{}{}{}
\acmYear{2018}
\copyrightyear{2018}

\begin{document}
\title{Verbal Disinhibition towards Robots is Associated with General Antisociality}

\author{Megan Strait, Virginia Contreras, \& Christian Duarte Vela}
\affiliation{%
  \institution{The University of Texas Rio Grande Valley}
  \streetaddress{1201 W University Drive}
  \city{Edinburg}
  \state{Texas}
  \postcode{78539}
}
\email{{megan.strait, virginia.contreras01, christian.duartevela01}@utrgv.edu}

\renewcommand{\shortauthors}{Strait et al.}

\begin{abstract}
The emergence of agentic technologies (e.g., robots) in increasingly public realms (e.g., social media) has revealed surprising antisocial tendencies in human-agent interactions.
In particular, there is growing indication of people's propensity to act \emph{aggressively} towards such systems -- \emph{without provocation and unabashedly so}.

Towards understanding whether this aggressive behavior is anomalous or whether it is associated with general antisocial tendencies in people's broader interactions, we examined people's verbal disinhibition towards two artificial agents.
Using Twitter as a corpus of free-form, unsupervised interactions, we identified $40$ independent Twitter users who tweeted abusively or non-abusively at one of two high-profile robots with Twitter accounts (TMI's Bina48 and Hanson Robotics' Sophia).
Analysis of 50 of each user's tweets most proximate to their tweet at the respective robot ($N=2,000$) shows people's aggression towards the robots to be associated with more frequent abuse in their general tweeting.
The findings thus suggest that disinhibition towards robots is not necessarily a pervasive tendency, but rather one driven by individual differences in antisociality.
Nevertheless, such unprovoked abuse highlights a need for attention to the reception of agentic technologies in society, as well as the necessity of corresponding capacities to recognize and respond to antisocial dynamics.
\end{abstract}



\maketitle

\section{Introduction}
Agentic technologies -- from disembodied AIs, to virtual agents, to highly humanlike robots -- are increasingly pervading public domains.
Disembodied AI assistants, such as Apple's Siri and Amazon's Alexa, have been available in consumer markets for nearly a decade.
Consumer- and enterprise-oriented robotic platforms, particularly those geared towards social engagement (e.g., Anki's Cozmo, Ugobe's Pleo, and Softbank's Pepper), have already achieved moderate commercial success\footnote{See, for example: \href{https://goo.gl/icsfYh}{goo.gl/icsfYh}.} and are beginning to emerge in public spaces around the globe\footnote{For example: \href{http://goo.gl/SwJYBY}{goo.gl/SwJYBY}, \href{https://goo.gl/UoX7oB}{goo.gl/UoX7oB}, and \href{https://goo.gl/x3v42f}{goo.gl/x3v42f}.}.
Furthermore, although virtual agents remain largely within academic settings, they show substantial potential for widespread public deployment across numerous industries including education (e.g., \cite{HewEtAl2010, MonahanEtAl2018}), medicine (e.g., doctor-patient communications \cite{BickmoreEtAl2007}), patient assistance \cite{BickmoreEtAl2009}), and therapy (e.g., evaluation \cite{LucasEtAl2015} and counseling \cite{LisettiEtAl2013}).

\begin{figure}[tb!]
	\includegraphics[width=.49\columnwidth]{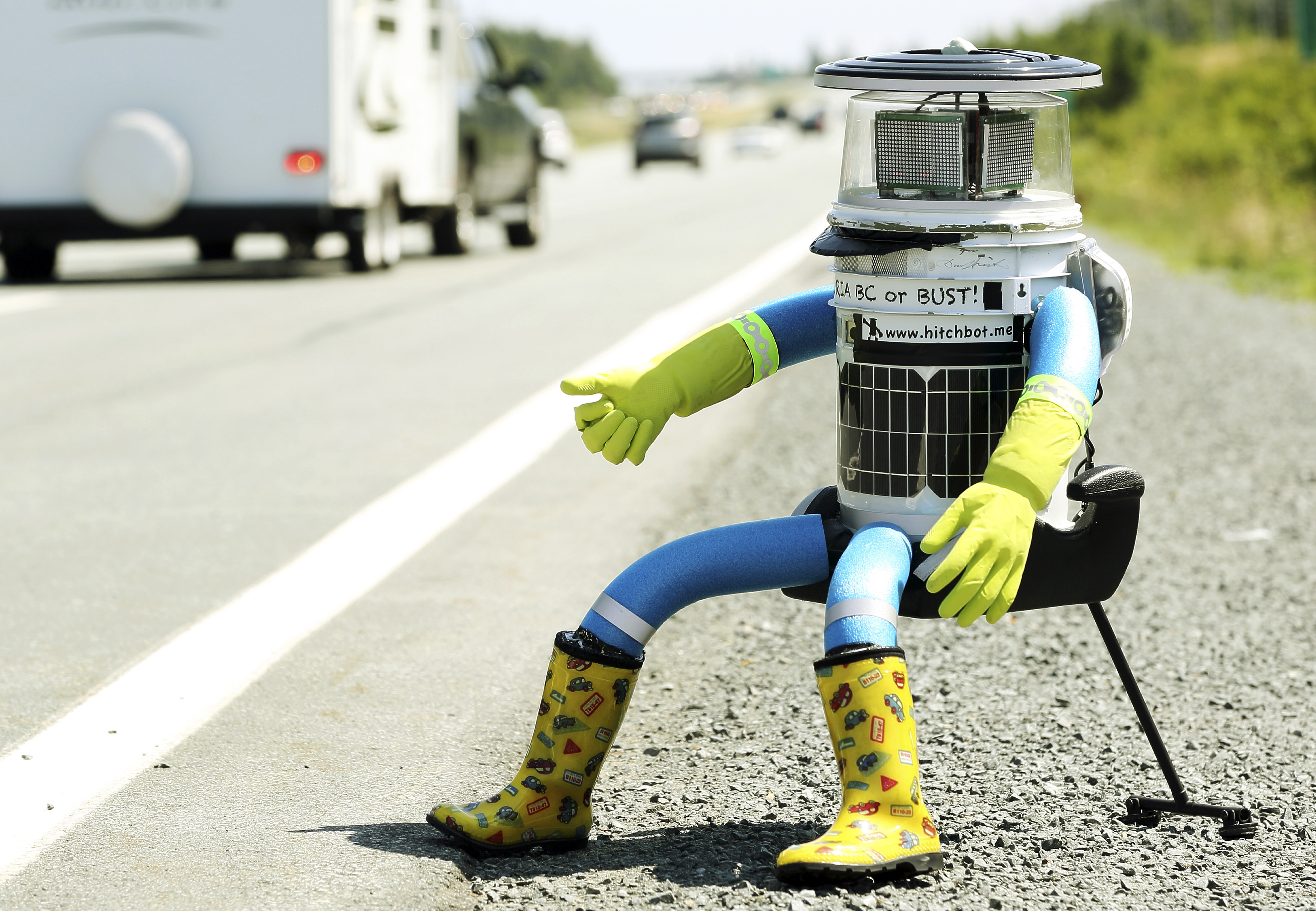}
	\includegraphics[width=.49\columnwidth]{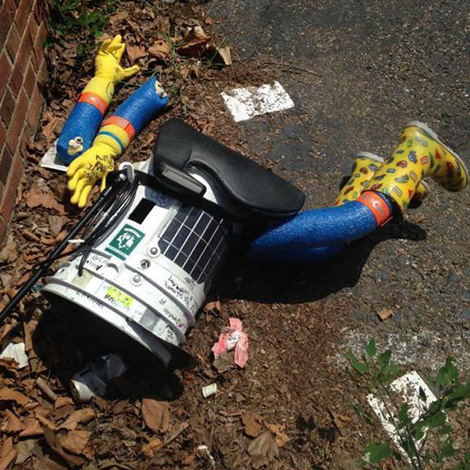}
	\caption{David Smith's and Frauke Zeller's hitchhiking robot, ``hitchBOT''. Shown is hitchBOT's original embodiment (left) and hitchBOT's decapitated and literally dis-armed remains after being vandalized in Philadelphia (right).}
	\label{HitchBOT}
\end{figure}

\subsection{Aggression in Human-Agent Interactions}
A natural result of their increased presence is that artificial agents are increasingly available for free-form, unsupervised interactions with the general public.
Observation of interactions in these more naturalistic settings have, in turn, brought to light people's apparent aggression toward agentic technologies.
For example:
\begin{itemize}
    \item \textbf{Ugobe's Pleo}: In 2007, DVICE released a video\footnote{\href{https://youtu.be/pQUCd4SbgM0}{https://youtu.be/pQUCd4SbgM0}} of a couple of staff members subjecting a Pleo robot to a series of abusive tests (including hitting the robot, smashing it against a table, and strangling it), which ultimately resulted in the robot's ``death''.
Though the tests were well-reasoned (to understand how Pleo acts in and responds to certain situations), the staffers' laughter throughout the abuse reflects a lack of empathy -- despite the robot embodying several responses intended to induce empathetic responding.
Moreover, the video's meta data, which shows it to have garnered several hundred explicit \emph{likes}, suggests that the staffers were not alone in their amusement.

\item \textbf{Smith and Zeller's hitchBOT}:
In 2014, researchers David Smith and Frauke Zeller launched a social experiment with their ``hitchBOT'' -- a robot designed to navigate and travel substantial distances by hitchhiking.
After two initial deployments (traveling across Canada, from Halifax to Victoria; as well as around Germany), hitchBOT was decapitated just two weeks into its deployment in the United States (see Figure~\ref{HitchBOT}).

\item \textbf{Microsoft's Tay}:
In 2016, Microsoft launched a similarly ill-fated social experiment -- deploying a chatbot (``Tay'') they had developed via Twitter.
Within 16 hours after its release, Tay, which was designed to learn from its interactions, morphed from its initial ``cheery teenage girl'' persona into a sexist, genocidal racist -- a direct result of the deluge of abuse that people directed at the bot.\footnote{https://goo.gl/nFEfS1}
\end{itemize}

Similar observations have appeared in academic discourse as well.
Verbal abuse comprises a substantial portion of people's commentary toward artificial agents, with observed frequencies ranging from 10\% (e.g., \cite{DeAngeliAndBrahnam2008}) to over 40\% (e.g., \cite{StraitEtAl2017}).
In interactions with agents that have physical embodiment, verbal abuses readily escalate to physical violence (including \emph{kicking}, \emph{punching}, and \emph{slapping}; \cite{BrscicEtAl2015, SalviniEtAl2010}).
Furthermore, the aggression occurs with or without supervision.
For example, in the supervised deployment of a virtual agent in educational settings, nearly 40\% of students were abusive toward the agent, employing, in particular, hypersexualizing and overtly dehumanizing commentary (e.g., \emph{``shut up u hore''}, \emph{``want to give me a blow job''}, \emph{``are you a lesbian?''}; \cite{VeletsianosEtAl2008}).

This deviation from socially normative behavior is not particularly surprising when considered alongside broader experimental research, which reflects a gap in the degree to which people empathize with artificial agents relative to the degree to which we empathize with other people (e.g., \cite{BartneckEtAl2005, RosenthalVonDerPuttenEtAl2013}).
Specifically, while there is ample evidence that people treat agentic technologies \emph{like} they do people (the ``media equation'' \cite{ReevesAndNass1996}), the treatment is \emph{not} equivalent.
For example, people's empathy towards a robot monotonically \emph{decreases} from androids (highly humanlike robots) to robots of more mechanomorphic appearances \cite{RiekEtAl2009}.
That is, the less human an agent seems, the less people empathize.
People also exhibit less empathy when observing a robot's (versus a person's) abuse \cite{RosenthalVonDerPuttenEtAl2013}, more readily engage in the abuse of a robot (versus of a person; \cite{BartneckEtAl2005}), and are generally unmoved by a robot's pleas for sympathy (e.g., \cite{BriggsEtAl2015, BrscicEtAl2015, JungEtAl2015, TanEtAl2018}).

\subsection{Implications \& Considerations}
These antisocial tendencies (aggression toward, and limited empathy for, agentic technologies) are especially problematic for two reasons in particular.
First, while aggression may not necessarily pose harm to a nonhuman target, aggression in the context of multi-party interactions negatively impacts bystanders who are witness to the abuse \cite{ZapfEtAl2011}.
Second, aggression toward humanlike robots -- which embody identity characteristics (e.g., gender) -- may facilitate subsequent aggression toward people who share identity characteristics with the abused robots.
For example, stereotypic abuse of a female-gendered robot (e.g., via sexualization) may reinforce stereotypes the aggressor has of women, resulting in greater expression of bias in subsequent interactions with women.

It is thus critical for artificial agents to be able to respond to manifestations of aggression if and when it arises.
To respond, however, requires that the agent has the capacity to recognize aggression.
And to accurately and reliably recognize aggression requires, first, identification of the relevant information channels and cues that communicate aggression.
To that end, recent work has identified a range of associated factors (e.g., the agent's gendering \cite{BrahnamAndDeAngeli2012}, racialization \cite{StraitEtAl2018}, and size \cite{LucasEtAl2016}).

Not all people, however, exhibit aggressive tendencies toward robots.
For example, deployment of a delivery robot in medical settings showed that while some staff treated the robot poorly and locked it away when they could, others treated the robot relatively well, using the robot to make their daily routine more efficient \cite{MutluAndForlizzi2008}.
Indeed, the majority do \emph{not} condone \cite{TanEtAl2018} nor exhibit aggression, with estimates as to the prevalence ranging from $8\%$ (explicit physical abuse; \cite{BrscicEtAl2015}) to just under $50\%$ (verbal aggression; \cite{StraitEtAl2017}).

\subsection{Present Work}
\begin{figure}[b!]
	\includegraphics[width=.49\columnwidth]{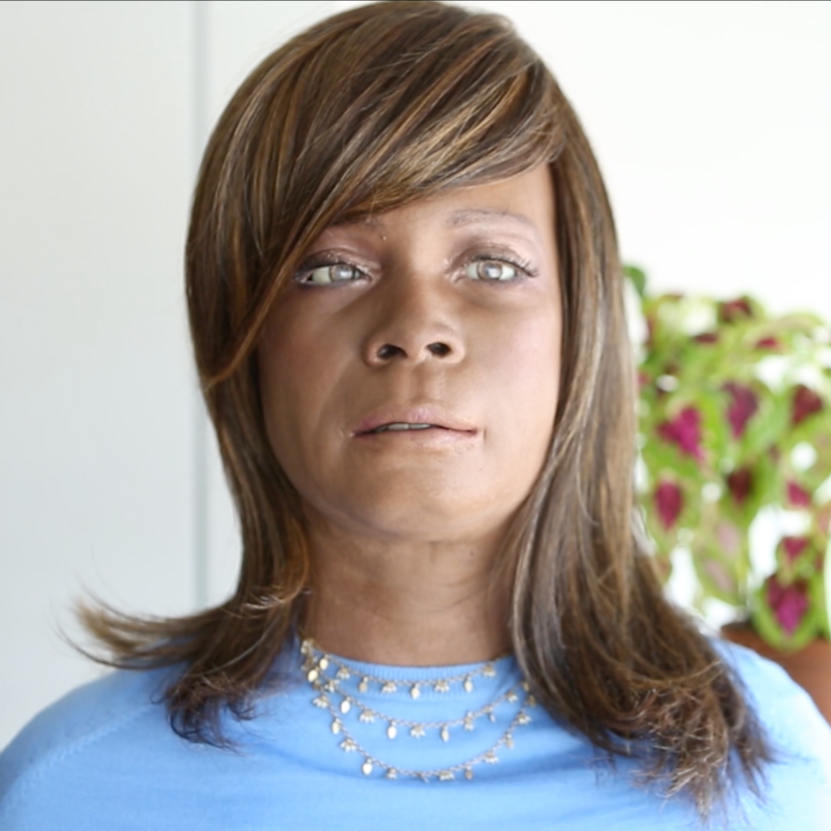}
	\includegraphics[width=.49\columnwidth]{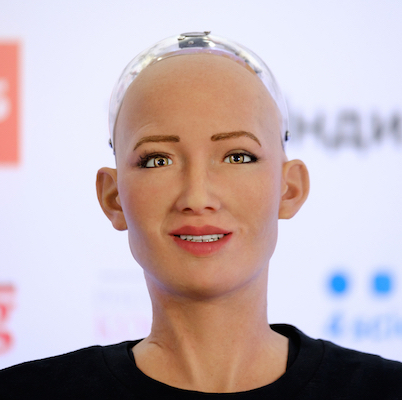}
	\caption{The two robots involved -- TMI's Bina48 (left) and Hanson Robotics' Sophia (right).}
	\label{Exemplars}
\end{figure}

Thus, towards better understanding differences amongst individual engagement in the aggressive treatment of agentic technologies, we examined the relationship between people's verbal abuse towards two robots -- TMI's Bina48 and Hanson Robotics' Sophia (see Figure~\ref{Exemplars}) -- and verbally aggressive tendencies in their interactions with other people.
Specifically, we sought to determine whether people's aggression toward the given robots is spontaneous or whether it is consistent with a broader pattern of aggression.
That is, is this a general phenomenon or does it align with individuals' prosociality (or rather, lack thereof).

To acquire data representative of more naturalistic (free-form, unsupervised) interactions with robots than what is available in controlled laboratory settings, we elected to scrape Twitter for commentary directed towards two robots with active accounts on Twitter.
For greater comparability to recent literature (e.g., \cite{SanchezramosEtAl2018, StraitEtAl2017, StraitEtAl2018}), we utilized \emph{robot} targets (versus other categories of agentic systems).
Towards mitigating associations stemming from any particular embodiment, we utilized \emph{two} targets.

From people's tweets at the two robots, we identified 40 distinct Twitter users (20 per robot) who tweeted abusively or non-abusively at the given robot.
We effected a quasi-manipulation of \textbf{user type} (two levels: \emph{abusive} versus \emph{non-abusive} towards robots) via identification of users (with $N_{abusive}=10$ and $N_{non-abusive}=10$ per robot).
We then scraped 50 of each user's tweets closest (in time) to their originating tweet ($N=2000$) in order to evaluate the association between aggression towards the robots and the prevalence of abuse in users' broader tweeting.

\section{Method}
We conducted an \textbf{online}, quasi-experimental evaluation ($N=40$) of the association between \textbf{aggression} in HRI and in human-human social dynamics.

\subsection{Design}
Given indications from existing literature that abusive human-agent interactions more frequently manifest in free-form, unsupervised contexts, we utilized Twitter (which hosts accounts for several publicized robot platforms such as Hanson Robotics' Sophia) as a source of similar interaction data.
Specifically, given greater disinhibition in online spaces \cite{Suler2004}, we expected to better capture abusive interactions that may not arise in more controlled contexts.
Furthermore, the interaction modality enables more naturalistc human-robot interactions than traditional laboratory settings \cite{SabanovicEtAl2006}, which may better capture the public's perceptions of emergent platforms.

Here we defined ``abusive'' as any content that is dehumanizing in nature.
Specifically, if a tweet contained content that was objectifying (including overt sexualization, \cite{MoradiAndHuang2008} and ambivalent and benevolent sexism \cite{GlickAndFiske1996}), racist (e.g., evocative of race-based stereotypes \cite{Allport1954}), generally offensive (e.g., such as calling the robot stupid \cite{BrscicEtAl2015}), and/or violent (verbally hostile or threatening of physical violence) towards the given agent -- it was coded as abusive.

\subsection{Manipulation}
We effected a quasi-manipulation of \textbf{user type} (\emph{abusive} versus {non-abusive}) via selection of the $40$ users ($20$ per robot, with $10$ abusive and $10$ non-abusive each).
To identify the $40$ users, we scraped all available Twitter mentions at Bina48 and Sophia (on March 22, 2018).
A total of $9,497$ tweets ($N_{Bina}=648$; $N_{Sophia}=8,849$) were returned -- a subset of which ($N_{Bina}=648$; $N_{Sophia}=1,000$) were then coded by a research assistant blind to the research questions on a single, binary dimension: whether a given tweet contains abusive content (1) or not (0).

\begin{table}[t!]
\caption{Source information from which the quasi-manipulation of \emph{user type} (abusive versus non-abusive) was effected. ``Mentions'', ``coded'', and ``retained'' refers to the number of tweets scraped, analyzed, and retained for selection of the 40 users.}
\label{TABLE:VIDEOS}
\centering
\begin{tabular}{cc ccc}
\hline\noalign{\smallskip}
\textbf{Source} 	& \textbf{Mentions}
				& \textbf{Coded}
				& \textbf{Retained}\\
\noalign{\smallskip}
\hline
\\
\textbf{\href{https://twitter.com/ibina48}{@iBina48}} 		& 648 & 648 	& $253$\\

\textbf{\href{https://twitter.com/realsophiarobot}{@RealSophiaRobot}}		& 8,849 & $1000$ & $631$\\
\\
\hline
\end{tabular}
\end{table}

A threshold of $1,000$ tweets for coding was set \emph{a priori} based on existing literature (using the lowest frequency of abuse reported in online contexts -- 10\% of commentary \cite{DeAngeliAndBrahnam2008}).
Although the expected proportion of abusive commentary (100) exceeds the number of abusive users needed (10), we set a higher threshold in anticipation of a lower frequency of abusive commentary (e.g., due to content moderation by the account managers) and loss of data (e.g., discarding of repeat tweets from the same user).
The criteria for retention were as follows:
\begin{itemize}
    \item \emph{Independence}: We aimed to identify independent users; thus, multiple tweets from a single user were excluded (except for one randomly selected tweet of the user's tweets). In addition, tweets which were replies to other users were excluded.

    \item \emph{Decipherability}: Any tweets that were indecipherable (e.g., due to lack of context) were excluded. For example, the tweet -- ``\emph{``\@iBina48: Cyber space'' \#pii2013}'' -- was excluded.
\end{itemize}
From the tweets remaining post-coding ($N_{Bina}=253$; $N_{Sophia}=631$), we randomly selected $20$ users ($10$ with an abusive and $10$ with a non-abusive tweet at the given robot) for each robot.

\subsection{Data Acquisition \& Annotation}
For each of the $40$ users selected (to effect the quasi-manipulation of \textbf{user type}), we scraped the user's 50 tweets most proximate to and centered around (i.e., 25 pre- and 25 post-) the user's originating tweet at one of the robots.
This scraping was completed between February 22 and March 02, 2018 and yielded a total of $2,000$ tweets for analysis.
Each of the $2,000$ tweets were coded on a binary dimension (0 or 1) for the presence of abusive content --  which was then used to compute an overall \textbf{frequency of abuse} for each of the $40$ users.
As verification of the coding reliability, a second coder independently coded 10\% of the $2,000$ tweets.
Calculation of Cohen's $\kappa$ confirmed high inter-rater reliability ($\kappa=.86$).

\section{Results}
Similar to rates reported in literature on verbal disinhibition towards chatbots (e.g., \cite{DeAngeliAndBrahnam2008}), the overall frequency of dehumanizing content across users comprised approximately 10\% of the Twitter-based interactions ($M=.09$, $SD=.12$).

To evaluate the association between \textbf{user type} (\emph{abusive} versus \emph{non-abusive} towards robots) and the frequency of abusive content in a user's general tweeting, we conducted an analysis of variance (ANOVA) with significance evaluated at a standard $\alpha$-level of $.05$.
Due to different racializations of the two robots (Bina48 is racialized as Black, Sophia is racialized as White), we included \textbf{robot racialization} as a covariate in the statistical model.

The results of the ANOVA showed a main effect of user type (\emph{abusive} versus \emph{non-abusive} towards robots) on the frequency of dehumanizing content in users' broader Twitter communications: $F=11.67$, $p<.01$, $\eta_{p}^{2}=.25$.
Specifically, the users identified in the coding process as abusive were much more frequently abusive in their general tweeting ($M=.15$, $SD=.15$) than were non-abusive users ($M=.03$, $SD=.05$; Cohen's $d=1.07$).
We additionally confirmed, via a post-hoc power analysis using the found effect size, that the study was adequately powered ($1-\beta=.9532$) to capture the given differences.

\section{Discussion}
\subsection{Summary of Findings}
The present study served as a preliminary investigation into individual differences in aggression towards agentic technologies.
Via an analysis of public tweet data, we found a significant association between people's antisociality and their abuse of two robots.

Given the methods used (wherein we evaluated the prevalence of aggression in each user's 50 surrounding tweets), there are two possible interpretations of this association:
(1) A person's aggression towards the robots is associated with an antisocial \emph{personality} (i.e., relatively unchanging demeanor).
(2) Or, it may be that a person's aggression towards the robots resulted during a period of general negative \emph{affect} (i.e., temporally-constrained aggression).

\subsection{Implications}
Assumming the first interpretation (aggression towards robots is associated with antisocial personality), then manifestations of aggression might be predicted by tracking of indicative personality characteristics and averted by proactive avoidance of interlocutors identified as generally antisocial.
Assuming the aggression resulted from negative affect, then tracking interlocutors' general affect (e.g., positive, neutral, or negative) may facilitate prediction of potential aggression.
In this case, manifestations of aggression might be mitigated via targeted intervention to regulate the aggressor's emotional state.

Assuming either interpretation, the findings indicate, in particular, that in addition to linguistic content analysis, construction and maintenance of models of interlocutors encountered may be important to the prediction and recognition of aggression in HAIs.
For example, if an interlocutor shows general aggressive tendencies, this may be a valuable heuristic toward deciding, subsequently, whether given data (e.g., linguistic utterance) is likely aggressive or not.
Or, if an interlocutor exhibits emotional agitation, recognition by the agent could cue an intervention such as an exercise in emotion regulation.

More generally, the findings underscore a need for respective agent capacities (to recognize and respond to aggression).
This is especially relevant in multi-party contexts, wherein aggressive treatment of an artificial agent may have broader impacts both on immediate bystanders and on subsequent interactions involving the aggressor (e.g., facilitation of the dehumanization of people sharing identity characteristics with the targeted agent).
However, responding to aggression requires, first, that the agent can reliably detect it when it manifests.
And the present findings indicate that detection may be significantly facilitated by modeling individual interlocutors in addition to explicit conversational content.

\subsection{Limitations \& Avenues for Future Research}
There are a number of limitations to the present study, which serve to highlight avenues for future research.
In particular, we conducted an online evaluation of the association between people's general degree of aggression and their aggression towards two robots.
However, more representative interaction settings (i.e., ecological validity), as well as broader sampling across platforms (i.e., more than two robots) and agent types (i.e., chatbots, virtual agents, and robots), is needed to understand how the findings extend to actual human-robot interactions and more generally, to human-agent interactions.
In addition, given the two possible interpretations to the present findings (association with \emph{personality} and/or \emph{affect}), further research is needed to determine which interpretation -- and corresponding, which approach to responding -- is appropriate (if not both).

\section{Conclusions}
Towards understanding individual differences in aggressive tendencies in human-agent interactions, we examined people's verbal disinhibition in their tweeting at two robots and in their broader interactions.
Using Twitter as a corpus of free-form, unsupervised interactions, we identified $40$ independent Twitter users who tweeted abusively or non-abusively at one of two robots with Twitter accounts (TMI's Bina48 and Hanson Robotics' Sophia).
Analysis of each user's 50 tweets most proximate to their tweet at the robot shows people's abuse of the robots aligns with more frequent abuse in their general tweeting.
The findings thus suggest that disinhibition towards robots is not necessarily a pervasive tendency, as it is significantly associated with general antisocial behavior.
While interpretation of the findings is constrained by methodological limitations, such unprovoked abuse nevertheless highlights a need for particular attention to the social capacities of agentic systems and suggests that maintenance of a user-specific model may facilitate predication and interpretation of aggression in HAIs.

\bibliographystyle{ACM-Reference-Format}
\bibliography{bibliography}

\end{document}